\documentclass{cupconf}
\usepackage{graphicx}
\usepackage{lscape}

\newcommand\lscaperule{\rule{540pt}{.5pt}\\[.5\baselineskip]}

\title[Optically observable zero-age main-sequence O stars]{Optically observable zero-age main-sequence O stars}

\author[N.~R.\ Walborn]{N\ls O\ls L\ls A\ls N\ns R.\ns W\ls A\ls L\ls B\ls O\ls R\ls N}
\affiliation{Space Telescope Science Institute, 3700 San Martin Drive, Baltimore, Maryland 21218, USA}

\begin{document}
\maketitle

\begin{abstract}
A list of 50 optically observable O~stars that are likely on or very
near the ZAMS is presented.  They have been selected on the basis of five
distinct criteria, although some of them exhibit more than one.  Three of
the criteria are spectroscopic (He~\textsc{ii}~$\lambda$4686 absorption stronger
than in normal luminosity class~V spectra, abnormally broad or strong 
Balmer lines, weak UV wind profiles for their spectral types), one is
environmental (association with dense, dusty nebular knots), and one is
photometric (derived absolute magnitudes fainter than class~V).  Very few
of these stars have been physically analyzed, and they have not been
considered in the current framework of early massive stellar evolution.
In particular, they may indicate that the earliest, embedded phases are
not as large a fraction of the main-sequence lifetimes as is currently
believed.  Detailed analyses of these objects will likely prove essential
to a complete understanding of the early evolution of massive stars.
\end{abstract}

\section{Introduction}

It is often stated that zero-age main-sequence (ZAMS) O~stars should not 
be and are not observed.  This view arises from at least three sources:
star-formation theory, which suggests that the embedded accretion (merger?)
phases constitute a significant fraction of the main-sequence lifetimes
of massive stars (2.5~Myr for the most massive); statistical studies of 
UCHII and IR objects relative to optically observed ones; and detailed
physical analyses of optical O-star samples that find very few on the ZAMS.
For instance, Repolust, Puls, \& Herrero (2004) analyzed 24 relatively
bright O~stars and found only one, HD~93128 in the Carina Nebula compact 
cluster Trumpler~14, on the ZAMS.  However, selection effects may be
contributing to this view.  If the optically observable near-ZAMS phase
of massive stars is relatively brief, it must be sought in very young
regions, which may be distant and/or extincted.  Also, it is possible 
that some IR objects are no longer embedded, but rather viewed along
unfavorable sightlines in galactic disks or through local, peripheral
remanent dust clouds.  For example, if we did not have such fortunate lines 
of sight toward NGC~3603 and 30~Doradus, we might be quite confused about
their evolutionary status (Walborn 2002).  Such must be the case for at
least some objects.

Over the past 35 years, this author and others have encountered
numerous optical O~stars that appear to be very young for various morphological 
reasons.  These results are scattered throughout the literature and have not 
been generally recognized by star-formation and evolutionary specialists, or
even by quantitative spectroscopists, whose analyses are essential for the
former.  It is hoped that this summary presentation will provide a useful
stimulus toward rectifying the omission.  It will be seen that many of
the objects in question are in the Magellanic Clouds.  But also, HD~93128
will reappear in the plot.

\section{Categories of candidate ZAMS O stars}

The current sample of 50 optically observable, likely ZAMS O~stars is
listed in Table~1, along with some normal standard stars for comparison.  
The ZAMS candidate list was\break
\begin{landscape}
\begin{table*}
\indexsize
\begin{center}
\lscaperule
\begin{minipage}{590pt}\centering
\begin{tabular}{@{}llrrcccll@{}}
\multicolumn{1}{c}{ID} &Sp Type &\multicolumn{1}{c}{$V$} &$B-V$ &$E(B-V)$ &${V_0}-M_V$ &$M_V$  &\multicolumn{1}{c}{Comment}  &\multicolumn{1}{c}{Reference}\\[4pt]
\multicolumn{9}{c}{\textit {Vz}}\\[1pt]
HD 64315         &O6 Vnz         &9.24  &0.25  &0.57  &13.2  &$-5.7$                      & &V.~Niemela priv.\ com (Fig.~3)\\
HD 64568         &O3 V((f*))     &9.39  &0.11  &0.43  &13.2  &$-5.1$  &also sublum             &\qquad "   (Fig.~3)\\
HD 92206B      &O6 V((f))       &9.16  &0.17  &0.49  &12.2   &...                      & &N.~Morrell priv.\ com\\
HD 93128         &O3.5 V((f+))  &8.77  &0.24  &0.56  &12.2  &$-5.1$  &also sublum          &Walborn 1973b, 1982, 1995\\
HD 93129B       &O3.5 V((f+))  &8.9   &0.22  &0.54  &12.2  &$-4.9$  &also sublum             &\qquad{"}\\
CPD$-58^{\circ}$2611 &O6 V((f))  &9.63  &0.28  &0.60  &12.2  &$-4.4$  &also sublum             &\qquad{"}\\
CPD$-58^{\circ}$2620 &O6.5 V((f))&9.27  &0.18  &0.50  &12.2  &$-4.4$  &also sublum             &\qquad{"}\\
HDE 303311       &O5 Vz          &9.05  &0.13  &0.45  &12.2  &$-4.5$  &also sublum          &V.~Niemela priv.\ com (Fig.~3)\\
FO 15  &O5.5 Vz       &12.05  &\multicolumn{1}{c}{...}  &1.21  &12.2  &$-4.2$  &also sublum, $R=4.15$  &Niemela et al.\ 2006 (Fig.~3)\\
HD 150135        &O6.5 Vz((f))   &6.89  &0.17  &0.49  &...    &...                     &  &Niemela \& Gamen 2005 (Fig.~3)\\
HD 152590        &O7.5 V         &8.42  &0.14  &0.46  &11.5  &$-4.5$  &also sublum          &Martins et al.\ 2005\\[3pt]
LH2$-$96           &O7.5 Vz       &14.95 &$-0.17$  &0.15  &18.6  &$-4.1$  &also sublum          &Parker et al.\ 2001\\
LH9$-$1486         &O6.5 Vz       &14.20 &$-0.21$  &0.11  &18.6  &$-4.7$  &also sublum          &Parker et al.\ 1992\\
LH10$-$3073        &O6.5 Vz       &14.71 &$-0.10$  &0.22  &18.6  &$-4.6$  &also sublum             &\qquad " (Fig.~2)\\
LH10$-$3102        &O7 Vz         &13.55 &$-0.10$  &0.22  &18.6  &$-5.7$                        &  &\qquad " (Fig.~2)\\
LH10$-$3126        &O6.5 Vz       &14.32  &0.00  &0.32  &18.6  &$-5.2$  &also Vb                 &\qquad " (Fig.~2)\\
LH10$-$3204        &O6$-$7 Vz       &14.02 &$-0.17$  &0.15  &18.6  &$-5.0$  &also sublum             &\qquad {"}\\
30 Dor$-$171       &O6$-$8 Vz       &15.67  &0.26  &0.58  &18.6  &$-4.7$                     &  &Walborn \& Blades 1997\\
30 Dor$-$341       &O8$-$9 Vz       &14.40 &$-0.03$  &0.28  &18.6  &$-5.0$                        &  &\qquad {"}\\
30 Dor$-$803       &O3$-$5 Vz       &15.61  &0.33  &0.65  &18.6  &$-4.9$  &also sublum             &\qquad {"}\\
30 Dor$-$1340      &O7 Vz         &14.94  &0.01  &0.33  &18.6  &$-4.6$                        &  &\qquad {"}\\
30 Dor$-$1643      &O3$-$5 Vz       &15.51  &0.15  &0.47  &18.6  &$-4.5$  &also sublum             &\qquad {"}\\
30 Dor$-$1892      &O8.5 Vz       &15.63  &0.23  &0.54  &18.6  &$-4.6$                  &        &\qquad {"}\\
30 Dor$-$2270      &O7 Vz         &15.31  &0.09  &0.41  &18.6  &$-4.5$  &also sublum          &Parker 1993\\[3pt]
NGC 346$-$113      &OC6 Vz        &14.93 &$-0.22$  &0.10  &19.1  &$-4.5$  &also sublum          &Walborn et al.\ 2000\\[4pt]
\multicolumn{9}{c}{\textit{Vb}}\\[1pt]
${\theta}^1$ Ori C  &O6 Vp var   &5.13  &0.00 &0.32    &\enspace8.3  &$-4.5$  &also wk wind,        &Morgan \& Keenan 1973\\
                             &                    &          &        &            &       &             & sublum, $R=5.5$?\\
${\theta}^2$ Ori A  &O9.5 Vp     &5.08 &$-0.11$ &0.19    &\enspace8.3  &$-4.0$  &$R=5.5?$\\
HD 92206C        &O8.5 Vp        &9.05  &0.15 &0.46   &12.2  &$-4.5$                 &      &Walborn 1982\\
Herschel 36   &O7.5 V(n)     &10.30  &\multicolumn{1}{c}{...}  &0.85   &10.9  &$-5.2$  &also knot, $R=5.39$    &Arias et al.\ 2006\\[4pt]
\multicolumn{9}{c}{\textit{Weak Winds}}\\[1pt]
HD 5005A         &O6.5 V((f))    &7.76  &\multicolumn{1}{c}{...}  &0.42   &...    &...  &from $b-y$           &Walborn et al.\ 1985\\
HD 42088         &O6.5 V         &7.55  &0.06 &0.38   &...    &...  &also Vz, ``knot''    &Martins et al.\ 2005\\
HD 54662         &O6.5 V         &6.21  &0.03 &0.35   &...    &...                 &      &Walborn et al.\ 1985\\
\end{tabular}
\end{minipage}
\caption{O-Type ZAMS Candidates}
\lscaperule
\end{center}
\end{table*}

\addtocounter{table}{-1}
\begin{table*}
\indexsize
\begin{center}
\lscaperule
\begin{minipage}{540pt}\centering
\begin{tabular}{@{}llrrccrll@{}}
\multicolumn{1}{c}{ID} &Sp Type &\multicolumn{1}{c}{$V$} &$B-V$ &$E(B-V)$ &${V_0}-M_V$ &$M_V$ &\multicolumn{1}{c}{Comment} &\multicolumn{1}{c}{Reference}\\[4pt]
\multicolumn{9}{c}{\textit{Knots}}\\[1pt]
N11A$-$7 &O3$-$6 V  &14.69 &\multicolumn{1}{c}{...} &0.19 &18.6 &$-4.5$ &$y$, also Vz, sublum &Heydari-Malayeri et al.\ 2001 (Fig.~2)\\
30 Dor$-$409A &O8.5 V &17.05 &\multicolumn{1}{c}{...} &0.56 &18.6 &$-3.3$ &WFPC2, also sublum &Walborn et al.\ 2002a, CHORIZOS\\
30 Dor$-$409B      &O9 V          &17.08  &\multicolumn{1}{c}{...}  &0.49   &18.6  &$-3.0$  &WFPC2, also sublum     &\qquad "\\
30 Dor$-$1201      &O9.5 V        &15.83  &\multicolumn{1}{c}{...}  &0.37   &18.6  &$-3.9$  &WFPC2                   &\qquad "\\
30 Dor$-$1222      &O9 V(n)p      &15.11  &\multicolumn{1}{c}{...}  &0.39   &18.6  &$-4.7$  &WFPC2                   &\qquad "\\
30 Dor$-$1429A     &O3-4 V        &15.88  &\multicolumn{1}{c}{...}  &0.55   &18.6  &$-4.4$  &WFPC2, also sublum      &\qquad "\\[3pt]
N81$-$1            &O6$-$8:  &14.38 &\multicolumn{1}{c}{...} &0.07   &19.1  &$-4.9$  &$y$, also wk wind    &Heydari-Malayeri et al.\ 2002\\
N81$-$2            &O6$-$8:         &14.87  &\multicolumn{1}{c}{...}  &0.06   &19.1  &$-4.4$  &$y$, also wk wind, sublum   &\qquad "\\
N81$-$3            &O6$-$8:         &16.10  &\multicolumn{1}{c}{...}  &0.10   &19.1  &$-3.3$  &$y$, also wk wind, sublum   &\qquad "\\
N81$-$11           &O6$-$8:         &15.74  &\multicolumn{1}{c}{...}  &0.07   &19.1  &$-3.6$  &$y$, also wk wind, sublum   &\qquad "\\[4pt]
\multicolumn{9}{c}{\textit{Subluminous (or $R>3$?)}}\\[1pt]
30 Dor$-$83        &O9$-$9.5 V      &15.44 &$-0.02$ &0.29   &18.6  &$-4.0$               &        &Walborn \& Blades 1997\\
30 Dor$-$324       &O7$-$8 V        &15.18  &0.04 &0.36   &18.6  &$-4.5$                    &      &\qquad "\\
30 Dor$-$466       &O9 V          &15.55 &$-0.11$ &0.20   &18.6  &$-3.6$                     &     &\qquad "\\
30 Dor$-$661       &O3$-$6 V        &15.03  &0.14 &0.46   &18.6 &$-5.0$                     &     &\qquad "\\
30 Dor$-$713       &O3$-$6 V        &14.61  &0.03 &0.35   &18.6  &$-5.0$                    &      &\qquad "\\
30 Dor$-$791       &O3$-$5 V        &15.84  &0.31 &0.63   &18.6  &$-4.6$                    &      &\qquad "\\
30 Dor$-$1035      &O3$-$6 V       &14.73 &$-0.05$ &0.27   &18.6  &$-4.7$                &      &\qquad "\\
30 Dor$-$1170      &O3$-$6 V        &15.94  &0.14 &0.46   &18.6  &$-4.0$                   &       &\qquad "\\[4pt]
\multicolumn{9}{c}{\textit{Normal V Comparisons}}\\[1pt]
HD 46149         &O8.5 V         &7.56  &0.17 &0.48   &10.5  &$-4.9$  &$R=4$                  &Walborn \& Fitzpatrick 1990\\
HD 46150         &O5 V((f))      &6.72\rlap{v} &0.13 &0.45   &10.5  &$-5.6$  &$R=4$                  &Walborn et al.\ 2002b\\
HD 46202         &O9 V           &8.17  &0.17 &0.48   &10.5  &$-4.2$  &$R=4$                 &Martins et al.\ 2005\\
HD 46223         &O4 V((f+))     &7.25  &0.22 &0.54   &10.5  &$-5.4$  &$R=4$                     &\qquad "\\
15 Mon           &O7 V((f))      &4.65 &$-0.25$ &0.07    &\enspace9.4  &$-5.0$             &          &Walborn \& Fitzpatrick 1990\\
HD 93027         &O9.5 V         &8.72 &$-0.02$ &0.28   &12.2  &$-4.3$         &                 &\qquad "\\
HD 93028         &O9 V           &8.36 &$-0.06$ &0.25   &12.2  &$-4.6$          &             &Martins et al.\ 2005\\
HD 93204         &O5 V((f))      &8.42  &0.10 &0.42   &12.2  &$-5.0$              &            &\qquad "\\
HDE 303308       &O4 V((f+))     &8.17  &0.13 &0.45   &12.2  &$-5.4$          &             &Walborn et al.\ 2002b\\[4pt]
\textit{Notes:}\\[1pt]
\multicolumn{9}{l}{$(B-V)_0$: O2$-$O7, $-0.32$; O8$-$O9, $-0.31$; O9.5, $-0.30$}\\ 
\multicolumn{9}{l}{$(b-y)_0=-0.15$; $E(B-V)=1.49E(b-y)$ (Heydari-Malayeri et~al.\ 2001, 2002)}\\
\multicolumn{9}{l}{$R=3$ unless otherwise noted}\\
\end{tabular}
\end{minipage}
\caption{\textit{Continued}}
\lscaperule
\end{center}
\end{table*}
\end{landscape}

\noindent complete to the author's knowledge as of 
May 2006, although further candidates are being discovered as of this
writing, e.g., in the SMC cluster NGC~346 by Evans et~al.\ (2006), and 
in a new LMC fiber survey by I.~Howarth (private communication).  
The sample is divided into five categories according to the distinct, 
principal discovery criteria, although many of them actually display 
more than one, as noted in the Comments.  Three of these criteria 
are spectroscopic, one is environmental, and one is photometric.  
Discovery and/or data references are included in the Table.  The five 
categories will now be discussed in turn.

\subsection{Vz spectra}
\begin{figure}
\centering
\includegraphics[width=.9\textwidth]{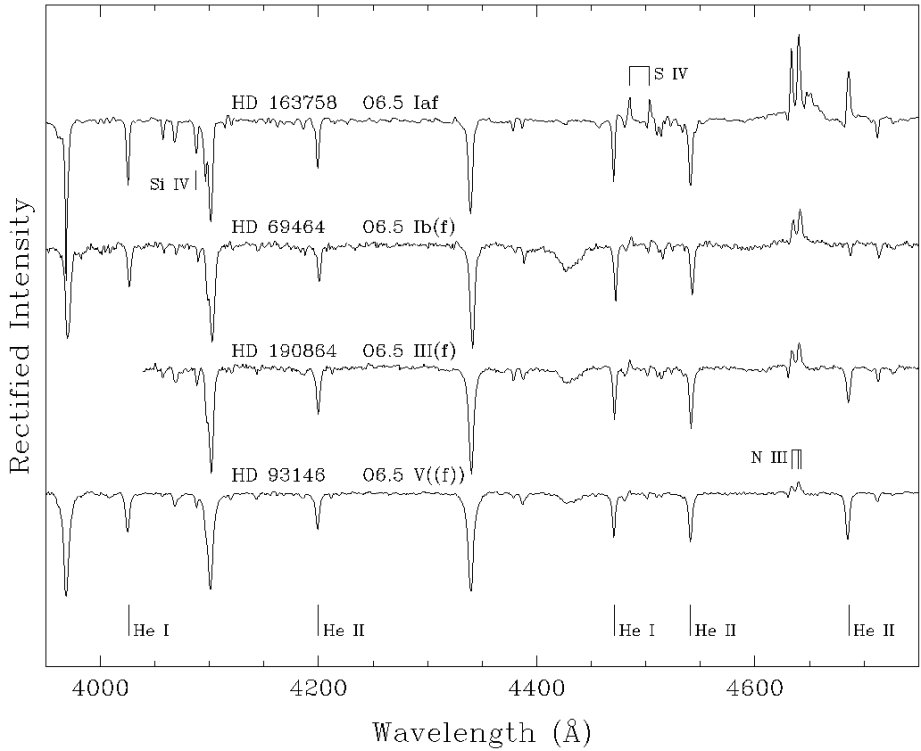}
\caption{A luminosity sequence at spectral type O6.5.  The rectified
spectrograms are separated by 0.4 continuum units.  The spectral lines
identified below are He~\textsc{i} $\lambda\lambda$4026, 4471 and He~\textsc{ii} 
$\lambda\lambda$4200, 4541, 4686.  N~\textsc{iii} $\lambda\lambda$4634-40-42 
emission is marked above HD~93146, likewise Si~\textsc{iv} $\lambda$4089 and S~\textsc{iv}  
$\lambda\lambda$4486-4504 (Werner \& Rauch 2001) in HD~163758.  Note 
the comparable strengths of the $\lambda\lambda$4541, 4686 absorptions 
in the class~V spectrum, and the weakening, then transition to emission
of the latter with increasing luminosity, while the N~\textsc{iii} emission
increases smoothly.  The Si~\textsc{iv} absorption has a positive luminosity
effect, which is more sensitive at later types.  Courtesy of Ian Howarth.}
\end{figure}

A luminosity classification for stars earlier than spectral type O9 was
introduced by Walborn (1971, 1972, 1973a).  It is based upon the
selective emission effects (Walborn 2001) in He~\textsc{ii}~$\lambda$4686 and the 
N~\textsc{iii} triplet $\lambda\lambda$4634-4640-4642, i.e., the Of phenomenon.  
These same lines display a \textit{negative} effect in absorption with
increasing luminosity in the MK O9--B0 classification, which was
hypothesized to be caused by filling in of the absorptions by the same 
emission effects producing the Of phenomenon, thus providing the basis 
for a luminosity classification at the earlier types.  A luminosity 
sequence at spectral type O6.5 in modern digital data is shown in Figure~1.  
As can be seen there, this He~\textsc{ii} line is a strong absorption feature in class~V, 
which then weakens, neutralizes, and finally comes into emission above the 
continuum in the Ia supergiant.  (Correlatively, the N~\textsc{iii}, already weakly 
in emission at class~V, increases in strength with increasing luminosity.)

\begin{figure}
\includegraphics[angle=90,width=\textwidth]{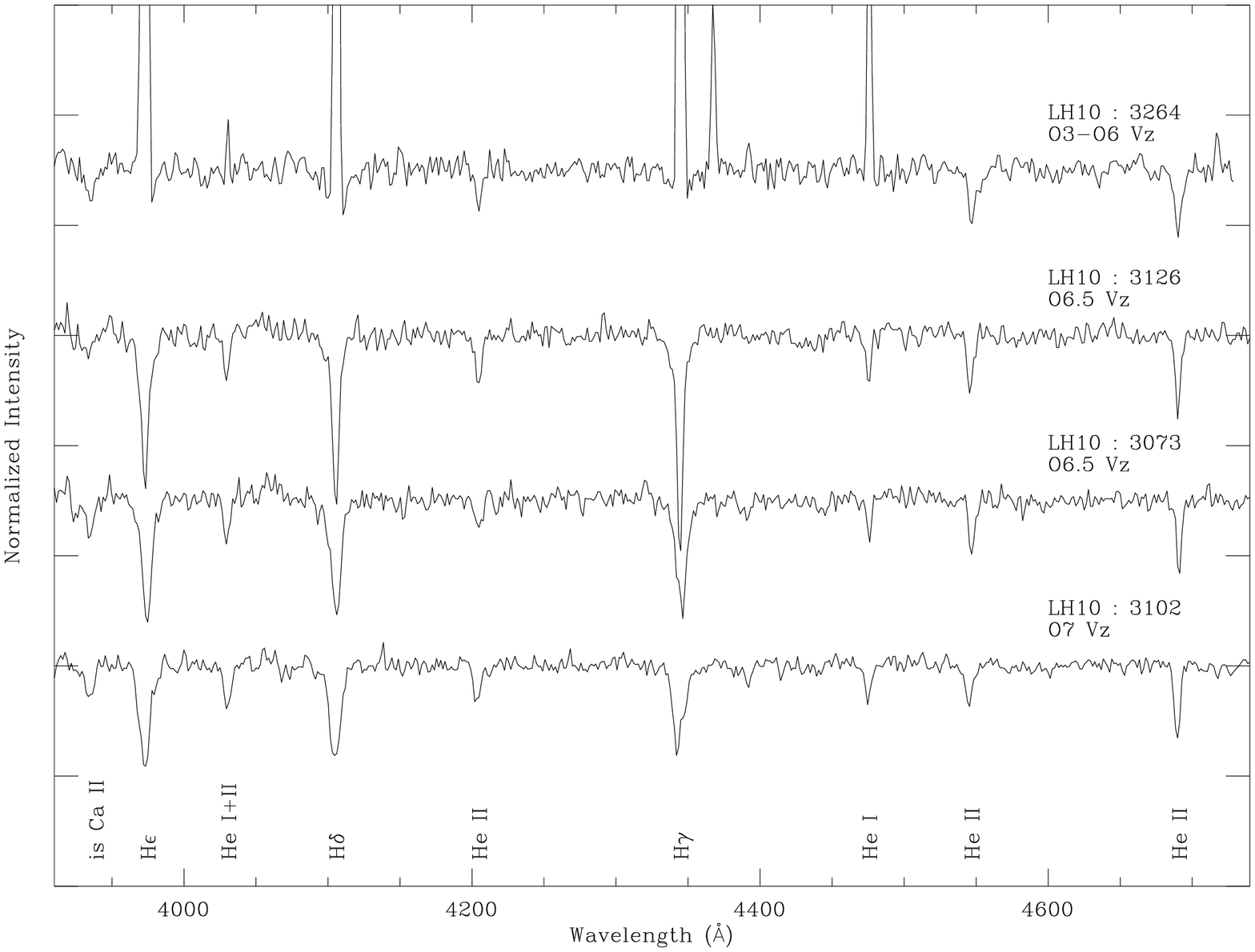}
\caption{Vz spectra in Lucke-Hodge 10/Henize N11 in the LMC (Walborn \&
Parker 1992).  The ordinate ticks are separated by 0.25 continuum units.
See the Fig.~1 caption for identifications of the He lines.  Note the 
increased strength of the He~\textsc{ii} $\lambda$4686 absorption relative to the 
other He lines in these spectra, and also the great strength of the Balmer
lines in LH10$-$3126.  LH10$-$3264 is in the dense nebular knot N11A, which
produces the very strong nebular emission lines.  Courtesy of Joel Parker. }
\end{figure}

Walborn (1973) first noted that in the Trumpler~14 O-dwarf spectra, the 
$\lambda$4686 absorption appeared stronger relative to the other He~lines
than in typical class~V spectra.  This very compact cluster in the Carina
Nebula appears to be very young; it contains the O2~If* prototype, likely 
pre-WN object HD~93129A (Walborn et~al.\ 2002b).  Penny et~al.\ (1993)
found that this cluster is approximately 550,000 years old, while Repolust
et~al.\ (2004) derived an age of 150,000 years for HD~93128.  Subsequently, 
even more extreme examples were found in the LMC giant H~\textsc{ii} regions 
30~Dor and N11 (Parker et~al.\ 1992; Walborn \& Parker 1992; Parker 1993; 
Walborn \& Blades 1997).  Some of the N11 spectra are reproduced in
Figure~2, and some new Galactic examples kindly provided for this
presentation by V.~Niemela and N.~Morrell are shown in Figure~3. 

\begin{figure}
\centering
\includegraphics[width=.9\textwidth]{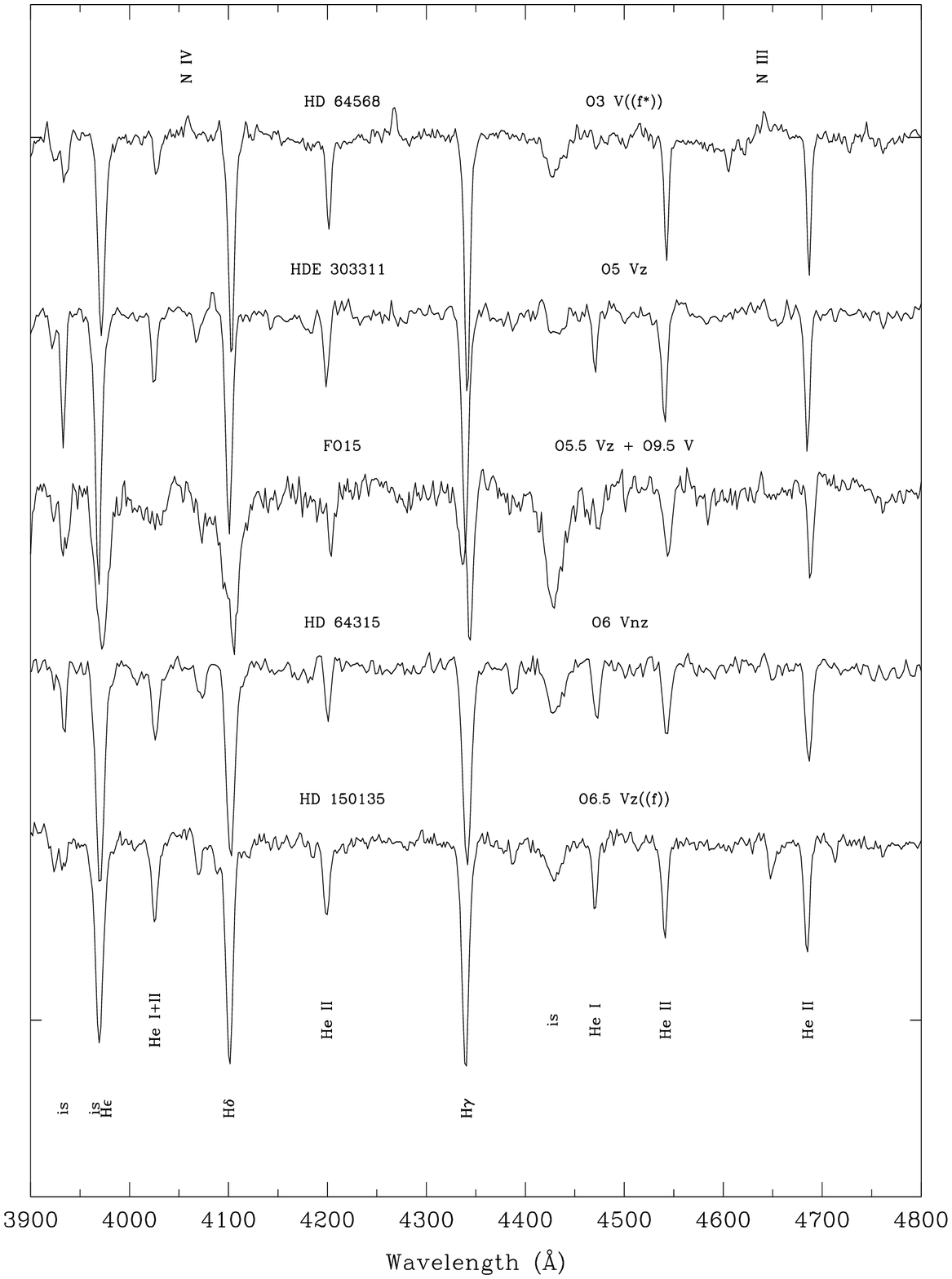}
\caption{New Galactic Vz spectra, rectified and separated by 0.2 continuum
units.  See the Fig.~1 caption for identifications of the stellar lines;
in addition, N~\textsc{iv} $\lambda$4058 emission is marked in HD~64568 here.
Courtesy of Virpi Niemela and Nidia Morrell.}
\end{figure}
  
A fairly obvious hypothesis is that the stronger $\lambda$4686 absorption
in these probably very young objects is an ``inverse'' Of effect, or more
precisely, that typical class~V spectra already have some emission
filling in that line while these objects have less, and hence may be less
luminous and less evolved, i.e., nearer to the ZAMS.  To denote that
hypothesis, the luminosity class notation Vz has been introduced.  Some
analytical support has been provided by the work of Venero, Cidale, \&
Ringuelet (2002), Martins et~al.\ (2005), and Mokiem et~al.\ (2006), but a
homogeneous analysis of the full sample in high-resolution data with 
state-of-the-art photospheric/wind models is essential to investigate 
whether the Vz stars as a class have systematically higher gravities and 
lower luminosities than class~V.  In Table~1, the three blocks of Vz stars 
correspond to the Galaxy, LMC, and SMC.

A misunderstanding of the Vz definition in late-O spectra should be
clarified here.  At early O~types, the He~\textsc{ii}~$\lambda$4686 absorption
should be stronger than He~\textsc{ii}~$\lambda$4541.  At type O7, He~\textsc{ii}~$\lambda$4541 
is equal to He~\textsc{i}~$\lambda$4471, so that $\lambda$4686 is
stronger than both in a Vz spectrum.  At later O~types, however,
$\lambda$4541 weakens more rapidly with advancing type than $\lambda$4686
in normal class~V spectra; thus, a late-O Vz spectrum must have 
$\lambda$4686 \textit{stronger than He~{\scriptsize I}~$\lambda$4471}.

\subsection{Vb spectra}

W.~Morgan frequently remarked on a peculiarity of the Orion Nebula Cluster 
OB~stars, namely broader Balmer lines than in normal class~V spectra, with
profiles that did not appear to be rotational; Morgan \& Keenan (1973)
reproduced photographic spectrograms of ${\theta}^1$~Ori~C and the O7~V
standard 15~Mon to illustrate the effect.  At some point, he introduced the 
notation Vb to denote such spectra, by analogy with Ib for less luminous
supergiants and IIIb in Keenan's subdivision of late-type giants.
(Abt 1979 cites a photographic atlas prepared with Morgan that appeared
in 1978 as the source of the new notation, but it does not appear there.)
Abt (1979) and Levato \& Malaroda (1981, 1982) presented further examples 
among B~and~A dwarf spectra in very young clusters.

As for the Vz category, the question arises whether the Vb phenomenon
might be caused by higher gravity and lower luminosity than in class~V,
since it is well known that the Balmer lines weaken with increasing
luminosity in OB~spectra due to the decreasing Stark effect.  A careful 
analysis of the five OB components of ${\theta}^1$ and ${\theta}^2$~Ori
has been presented by Sim\'on-D\'{\i}az et~al. (2006), including a
thorough investigation of line-broadening mechanisms and comparison with
standard objects (including 15~Mon) analyzed with the same techniques.  
Interestingly, they find that H and He lines in the Orion stars tend to be 
broader than in the best fitting models, most systematically in the B0.5~V 
spectrum of ${\theta}^1$~Ori~D.  However, they derive similar gravities to 
those of the comparison stars and find that the Orion stars are somewhat off 
the ZAMS, although uncertainties in the latter result remain because of the
extinction law and, as they point out, the effect of initial rotational
velocities on the location of the ZAMS.  In Table~1, a value of $R=4.25$ 
(average of 3.0 and 5.5) has been used to calculate the absolute magnitudes 
of the Trapezium stars, because the actual value in the Orion Nebula remains 
uncertain (Robberto et~al.\ 2004); thus their subluminosity is uncertain as 
well. 

${\theta}^1$~Ori~C is now known as the first O-type magnetic oblique rotator 
(Donati et~al.\ 2002; Smith \& Fullerton 2005; Gagn\'e et~al.\ 2005; Wade 
et~al.\ 2006) and an extreme spectrum variable, including its abnormally 
weak UV~wind profiles (Walborn \& Nichols 1994; Stahl et~al.\ 1996).  
Sim\'on-D\'{\i}az et~al. (2006) provide a detailed discussion of the
effects of the variability on the quantitative analysis.  Moreover, there 
is now evidence that the spectrum of 15~Mon may be variable as well;
Sim\'on-D\'{\i}az et~al. also discuss the effects of its spectroscopic
companion in that connection.  Further quantitative analyses incorporating 
these complications are indicated to ascertain the nature of the Vb Balmer 
profiles and the physical origin of the peculiarity.  

\subsection{Weak winds}

It is now well known that the ultraviolet stellar-wind profiles in O-type
spectra display strong correlations with the optical spectral types, including 
an increase in strength with increasing luminosity (Walborn, Nichols-Bohlin, 
\& Panek 1985).  That study also detected four main-sequence stars
with abnormally weak wind profiles for their spectral types, including 
${\theta}^1$~Ori~C already discussed in the previous section.  HD~5005A
in the young cluster NGC~281 is also located in a trapezium system, while 
HD~42088 ionizes the small, dusty H~\textsc{ii} region NGC~2175, which might appear
knot-like (cf.\ next section) if it were at the distance of the Magellanic
Clouds.  Again, the inverse behavior of these weak wind profiles with respect 
to normal spectra suggests that these stars may be less luminous and less 
evolved than typical class~V objects, but quantitative confirmation of this 
hypothesis is so far lacking.  Martins et~al.\ (2005) found a relatively
weak wind for HD~42088, but a normal HRD location; however, the distance
of this star is highly uncertain.  (It should be noted that most of the stars
in Table~1 have not yet been observed in the far UV.)

\subsection{Nebular knots}

Walborn \& Blades (1987) reported the detection of O-type spectra in two
dense nebular knots within the 30~Doradus Nebula in the LMC, suggesting that 
they might correspond to very young objects just emerging from their natal
cocoons.  Subsequent work, both from the ground and with \textit{HST}, has amply 
confirmed that suggestion and revealed additional examples in 30~Dor 
(Walborn \& Blades 1997, Walborn et~al.\ 1999, 2002a); several have been
resolved into multiple (trapezium) stellar systems by \textit{HST}.  The very strong
nebular emission lines in these objects obliterate the stellar He~\textsc{i}
absorption and thus preclude accurate classification, as only the He~\textsc{ii}
can be seen, leading to the O3-6~V type (there is no uncertainty in the
luminosity class, which depends on only He~\textsc{ii} $\lambda$4686); in fact, 
several examples re-observed spectroscopically with \textit{HST}, its very high
spatial resolution suppressing the nebular background, turn out to have
even later O~types than that range previously assigned from the ground
(Table~1).  The reddening and absolute magnitudes of these stars have
kindly been derived by L.~\'Ubeda from the WFPC2 photometry using the
CHORIZOS code (Ma\'{\i}z-Apell\'aniz 2004); a normal reddening law had to
be adopted because of the limited wavelength coverage.  

LH10$-$3264 (Parker et~al.\ 1992, Walborn \& Parker 1992; Fig.~3 here), 
which is also Vz, is another interesting case in the LMC compact H~\textsc{ii} 
region N11A; it has also been resolved into a multiple system with \textit{HST} by
Heydari-Malayeri et~al.\ (2001), and the entry in Table~1 (N11A-7) 
corresponds to the brightest component.  N81 is a similar object in the 
SMC, investigated with \textit{HST} by Heydari-Malayeri et~al.\ (2002); the 
weakness of the stellar-wind profiles in these stars may be caused by a
combination of extreme youth and the SMC metal deficiency.       

\subsection{Subluminosity}

In the spectroscopic study of the 30~Doradus stellar populations by
Walborn \& Blades (1997), the derived absolute magnitudes for a number of 
the O~V stars fall below the calibration of that luminosity class (cf.\ 
their Fig.~3).  Those that are not also classified as Vz are listed separately 
in Table~1.  It is reasonable to hypothesize that they may be nearer to or 
on the ZAMS.  However, a basic uncertainty, which also applies to some 
subluminous cases discussed in the previous sections, must be recognized: 
because of the lack of further information, the absolute magnitudes have
to be derived with a normal value of $R=3$.  For a fixed distance modulus
and variable extinction, a larger value of $R$ would yield brighter absolute 
magnitudes.  To resolve this issue, photometric observations covering a
wider wavelength range (ideally, UV through IR) must be obtained to
support the derivation of reddening laws toward each individual star.
Indeed, there is evidence that the reddening law may vary among different
lines of sight within a complex H~\textsc{ii} region environment, because of spatially 
diverse mixtures of dust grain properties.  In any event, large values
of $R$ are associated with large dust grains and very young objects, so
these stars may be near the ZAMS in either case, but again, further
observations and quantitative analyses are required to definitively establish
their physical status.

\section{Discussion}

All of the absolute magnitudes from Table~1 are plotted against the
spectral types in Figure~4, where they are also compared with the
Schaller et~al.\ (1992) ZAMS and isochrones, as well as the luminosity
class~V calibration (Walborn 1973).  The different categories of ZAMS 
candidates are distinguished with different symbols.  Subject to the 
uncertainties already discussed, the plot provides preliminary support for 
the present hypothesis: most of them fall below the class~V calibration and 
near the ZAMS.  Note that HD~93128 lies adjacent to the ZAMS, in agreement 
with the detailed analysis of Repolust et~al.\ (2004).  The normal class~V 
stars cluster about the calibration, as well they should since they 
contributed to its derivation.  Interestingly, the figure shows that a typical 
O3~V star is 1~Myr old, O5~V 2~Myr, O6--7~V 3~Myr, and O8--9~V perhaps 4--5~Myr.  
Less luminous stars at a given type must be younger.  It is also clear that 
unresolved multiple systems move points upward in the diagram (recall that 
0\mbox{$.\!\!^{\prime\prime}$}1 corresponds to 5000~AU at the LMC), while underestimated values 
of $R$ move them downward. 

\renewcommand{\baselinestretch}{1.1}
\begin{figure}
\centering
\includegraphics[width=3.5in]{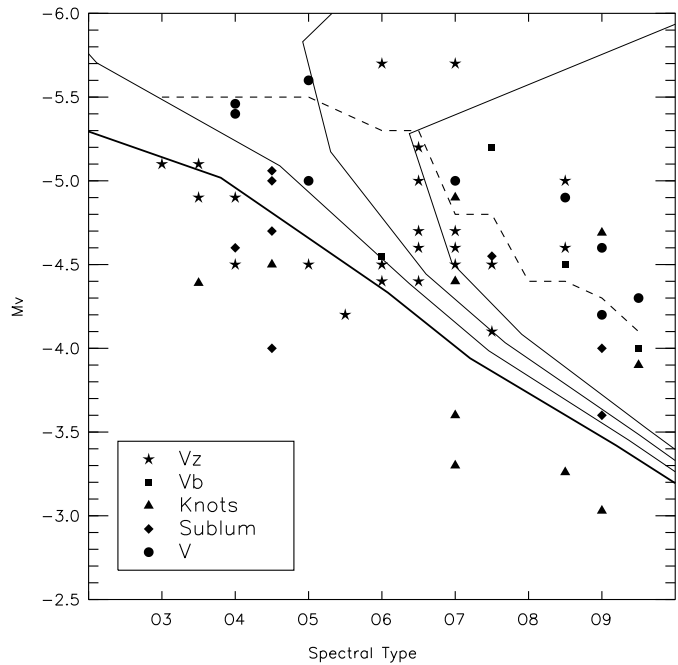}
\caption{Derived absolute visual magnitudes for candidate ZAMS and normal
O~stars plotted against their spectral types.  The bold solid line is the
ZAMS from Schaller et~al.\ (1992), while the lighter solid lines are the
corresponding isochrones for 1, 2, 3~Myr.  The dashed line is the
luminosity class~V calibration from Walborn (1973a).  The normal stars
fall near the latter, while most of the ZAMS candidates lie below it.
Courtesy of Leonardo \'Ubeda.}
\end{figure}
\renewcommand{\baselinestretch}{1}

The present sample of candidate ZAMS stars is optimum for followup with
high-resolution spectroscopy and state-of-the-art atmospheric/wind
analysis.  Beyond the basic question of their ZAMS status or otherwise,
it may be hoped that such study will elucidate any sequential
relationships among the different categories, thereby advancing our
detailed understanding of early massive stellar evolution.

\begin{acknowledgements}
My sincere thanks to the following colleagues and STScI staff who
contributed data, plots, and/or technical support for this presentation
and publication:  Rodolfo Barb\'a, Howard Bond, Artemio Herrero,   
Mohammad Heydari-Malayeri, Ian Howarth, Jes\'us Ma\'{\i}z-Apell\'aniz, 
Greg Masci, Nidia Morrell, Virpi Niemela, Joel Parker, Calvin Tullos, 
Leonardo \'Ubeda, and Skip Westphal.
\end{acknowledgements}

\end{document}